%% file: main.tex
\documentclass[conference]{IEEEtran}

\IEEEoverridecommandlockouts

\ifCLASSINFOpdf
  \usepackage[pdftex]{graphicx}
  \DeclareGraphicsExtensions{.pdf,.jpeg,.png}
\else
  \usepackage[dvips]{graphicx}
  \DeclareGraphicsExtensions{.eps}
\fi

\ifCLASSOPTIONcompsoc
  \usepackage[caption=false,font=normalsize,labelfont=sf,textfont=sf]{subfig}
\else
  \usepackage[caption=false,font=footnotesize]{subfig}
\fi

\usepackage[colorlinks,urlcolor=blue,linkcolor=blue,citecolor=blue]{hyperref}

\usepackage{paralist}

\newcommand{\mxfigfirst}[1]{{\bf Figure~\ref{fig:#1}}}
\newcommand{\mxfigref}[1]{Figure~\ref{fig:#1}}

\begin{document}

\title{Research Challenges for Heterogeneous CPS Design
\thanks{This is a pre-publication version of a paper that has been
accepted for publication in IEEE Computer.
The official/final version of the paper
will be posted on IEEE Xplore.}}

\author{\IEEEauthorblockN{Shuvra S.~Bhattacharyya\IEEEauthorrefmark{1},
                          Marilyn C.~Wolf\IEEEauthorrefmark{2}}
\IEEEauthorblockA{\IEEEauthorrefmark{1}University of Maryland, College Park}
\IEEEauthorblockA{\IEEEauthorrefmark{2}University of Nebraska, Lincoln} 
}

\maketitle

\begin{abstract}
\input{abstract}
\end{abstract}

\section{INTRODUCTION}
\input{intro}

\section{INTEROPERABILITY IN CPS}
\label{sec:interoperability}
\input{interoperability}

\section{COMPACT SYSTEM-LEVEL MODELS}
\label{sec:csm}
\input{csm}

\section{MODELING EXAMPLE}
\label{sec:example}
\input{example}

\section{OUTLOOK}
\input{conclusion}

\section{ACKNOWLEDGMENT}
\input{ack}

\bibliographystyle{IEEEtran} 
\bibliography{refs}

\end{document}

%% file: abstract.tex
Heterogeneous computing is widely used at all levels of computing from data
center to edge due to its power/performance characteristics. However,
heterogeneity presents challenges.  Interoperability---the management of
workloads across heterogeneous resources---requires more careful design than is
the case for homogeneous platforms. Cyber-physical systems present additional
challenges. This article considers research challenges in heterogeneous CPS
design, including interoperability, physical modeling, models of computation,
self-awareness and adaptation, architecture, and scheduling.

%% file: intro.tex
Heterogeneous computing offers the potential to streamline
execution of key tasks for processing, sensing, actuation, and communication
using devices that are better suited to those tasks than architectures composed
from collections of identical devices. This potential is of great utility for
cyber-physical systems (CPSs), where constraints on energy consumption, cost
and real-time performance often motivate the investigation of highly
streamlined solutions.  However, increased use of heterogeneity leads to complex
challenges and important needs associated with interoperability and model-based
design in CPSs.  This paper outlines challenges in heterogeneous CPS design,
and motivates the need for approaches to system-level design that are based on
complementary collections of compact system-level models.

%% file: interoperability.tex
Interoperability has been studied in many different forms in the context of
heterogeneous computing and CPS.  In this section, we review a small
sampling of 
representative directions of investigation.  A major direction of the recent
emphasis in heterogeneous computing has focused on interoperability in the
context of cloud computing (e.g., see~\cite{ranj2014x1}). Interoperability in
this context involves both the management of application workloads across
heterogeneous collections of resources associated with a given cloud computing
service provider as well as the deployment of workloads across resources of
different providers.

Givehchi {\em et al.}~investigate interoperability challenges in industrial
cyber-physical systems involving the networked management
of data from heterogeneous field devices, such as
I/O devices, sensors, and actuators~\cite{give2017x1}.
They introduce an interoperability layer
for connecting the physical and cyber layers
in networked factory systems in such a way
that legacy devices can be integrated without modification.

G\"{u}rd\"{u}r  {\em et al.}~present a survey of methods for assessing
interoperability in tool chains for CPS~\cite{gurd2016x1}. They identify
numerous assessment models and focus on fourteen of the most popular models,
which have been introduced over a period spanning 1980--2007.  Their
investigation found that most of the assessment models focus on isolated types
of interoperability, and rely on complex metrics, which limits their usability
in practical CPS design contexts.

In this paper, we discuss approaches for enhancing
interoperability and heterogeneous CPS design
based on the use of complementary
modeling strategies, which abstract different
concerns in the design process through well-defined, 
formal modeling concepts. We emphasize the 
diversity of different design concerns that
may be modeled in this way, and the need 
for compactness in the models that are employed.

%% file: csm.tex
Raising the level of abstraction in design processes for
CPS can facilitate interoperability by making it easier
to reason about the behavior of subsystems in a design and interactions 
between them. However, due to the multi-faceted nature of CPS
system design, no single abstraction or small set of abstractions
is adequate for design of all systems. Instead, the abstractions to
employ must be selected and applied in complementary ways that are well
matched to the targeted class of applications, and the 
objectives and constraints that are involved in their design.

Given the complexity of modern CPS systems, the size of
the models in the employed abstractions is an important
consideration in their formulation or selection.
The transition from assembly language to high-level languages
such as FORTRAN or C, which began many decades ago,
can be considered as an increase in the level
of abstraction. However, modern CPS systems involve
hundreds of thousands to tens of millions of lines of 
high-level language code or more. The compactness 
of the models that are involved in the abstractions
becomes an important concern to facilitate
human understanding and tractable analysis of the models.

Strategic application of compact models is important, for example, in the
paradigm of dynamic, data driven applications systems (DDDAS), where an
executing model of an application is integrated into a feedback loop with
instrumentation processes that supply data to the model~\cite{blas2018x2}.
Accurate, compact models are useful for real-time adaptation of DDDAS models
based on dynamic changes in the data acquired from instrumentation, and
conversely, for control of the instrumentation processes by the executing
models.

The motivations above for diverse and compact abstractions
leads us to advocate the concept of {\em compact system-level models}
as a central concept in the design and implementation of CPSs.
Many different types of models are relevant
to CPS design. Some prominent examples include the following.

\begin{asparaitem}
\item Models of physical phenomena~\cite{wolf2016x2}.  Computing is a physical
act: it takes time and energy; the reliability of the result depends on the
physics of the computing system.  Taking all these physical phenomena into
account in multi-billion-transistor systems is extremely challenging.

\item Models of computation. A model of computation defines how an
interconnected set of components interact to perform computation. A few
examples of important classes of models of computation include dataflow models,
state machines, and discrete event models.  Models of computation may impose
restrictions on how components are defined or interact that make important
analysis or optimization problems become tractable (e.g,
see~\cite{eker2003x1}). In contrast, fundamental analysis problems, such as
whether a program halts or has bounded memory requirements, are undecidable
in conventional programming languages for general-purpose computing.
Models of computation contribute to modeling compactness
by abstracting implementation details of
individual functional components and their coordination.

\item Models of self-awareness and adaptation~\cite{dutt2016x1}. Stochastic
models provide systems with compact, run-time-ready models that they can use to
estimate their own state.  Training allows us to capture complex models, so
long as we have sufficient training data.  Once trained, those models can be
evaluated much more efficiently on the platform.  Their results allow the
system to reflect on its own power and thermal behavior.  Managing power and
thermal behavior is critical to maintaining system longevity.

\item Models of architecture.
While models of computation focus on capturing the algorithmic
behavior of application systems, models of architecture 
provide compact abstractions of the hardware on which the
algorithms are mapped~\cite{pelc2018x2}. Models of architecture
are formulated to enable efficient, reproducible
estimation of nonfunctional costs associated with executing applications
that are described in terms of a given model of computation.
These costs include important metrics for efficiency
evaluation, such as latency, throughput, memory requirements,
and energy consumption.
A key concept in the formulation of models of architecture is the 
decomposition of application execution into
quantized units of communication and computation,
and the estimation of costs in terms of these abstract units.
Models of architecture are more constrained and operate
at a higher level of abstraction compared to hardware
description languages, such as Verilog and VHDL.

\item Scheduling models. Scheduling is an important aspect
of implementation that is abstracted away by models
of computation. Scheduling involves the assignment of
computational tasks to processing resources and the
ordering of tasks that share common resources. 
Scheduling often has major impact on metrics for 
efficiency evaluation, including the ones listed above.
Model-based scheduling representations provide formal,
platform-independent approaches for representing,
reasoning about, and transforming schedules~\cite{bhat2018x4}.
\end{asparaitem}

A design methodology based on compact system-level models for CPS involves the
selection of such models, and the definition of how representations and design
tools associated with these models are cooperatively applied in system design
processes.  While there are trade-offs between model complexity and accuracy
that may be involved in the models that are employed, restricting attention to
only the highest fidelity models may severely limit the extent of the design
space that can be investigated.

%% file: example.tex
\begin{figure*}
\centerline{\includegraphics[width=40pc]
{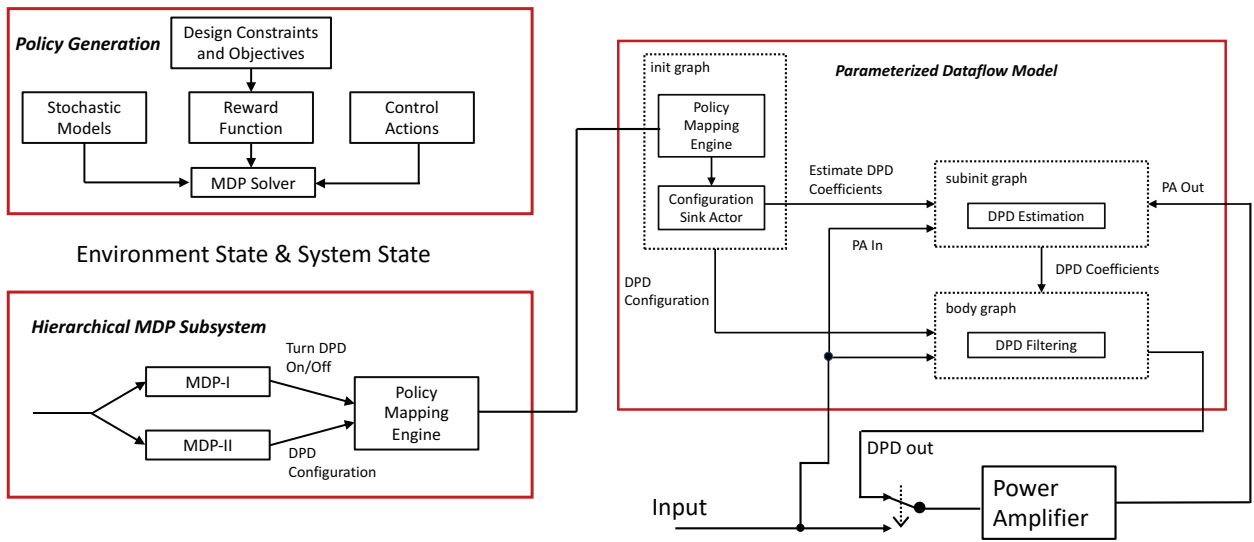}}
\caption{An illustration of the MADS framework (adapted from~\cite{li2019x8}).}
\label{fig:mads}
\end{figure*}

An example of a complex subsystem design using multiple forms of compact
system-level models is the the MDP framework for Adaptive DPD (digital
predistortion) Systems (MADS)~\cite{li2019x8}. DPD is a type of algorithm that
is used to counteract nonlinearities in power amplifiers (PAs) to improve the
quality of wireless communications signals ({\em e.g.}, see
~\cite{antt2010x1}).  The design and configuration of DPD systems involves
complex trade-offs among signal quality, energy efficiency and real-time
performance.  The MADS framework is demonstrated by mapping it into an
optimized implementation on a CPU/GPU platform. The model-based design of the
MADS framework is illustrated in \mxfigfirst{mads}.

The MADS framework illustrates an approach to several of the challenges
associated with heterogeneous CPS design discussed in this paper.  MADS
applies a model of the physics involved in a communications transmitter to
define, simulate and fine-tune the core predistortion algorithm that is
employed. A Markov decision process (MDP) is employed in MADS as a model that
provides self-awareness and adaptation capabilities. MDPs are probabilistic
models that are used to derive adaptation policies in uncertain environments.
In particular, MDPs are used in the context of environments that are
characterized using memoryless probability distributions --- that is, the
distribution of the next state is dependent only on the current state, and not
on the trajectory of prior states that led to the current state. In MADS,
MDP-based DPD architecture adaptation is performed with the objective of
jointly optimizing signal quality, system throughput, and power consumption.

In general, MDP models can become large and unwieldy to employ in complex
applications. To help ensure compactness of the MDP model that is employed, a
hierarchical MDP~\cite{jons2006x1} structure is designed, as illustrated in the
lower left part of \mxfigref{mads}.  

Parameterized dataflow~\cite{bhat2001x2}
is used in MADS as a model of computation to represent the algorithms employed
for adaptation and DPD operation, and model their interactions. 
In parameterized dataflow, the design for
a signal processing system is decomposed into three cooperating
dataflow graphs, called the init graph, subinit graph, and body graph
(see \mxfigref{mads}).
The body graph represents the core signal processing functionality,
while the init and subinit graphs represent functionality for
dynamic manipulation of parameters in the body graph. The 
init and subinit graphs differ in the frequency with which the
associated parameter adaptation operations are carried out, with
subinit graph operations being more frequent~\cite{bhat2001x2}.
In MADS, the parameterized dataflow model is used as a starting
point to map the MDP-equipped adaptive system into a CPU/GPU implementation.

For more details on the MADS framework, including the
different design components illustrated in \mxfigref{mads}, we
refer the reader to the presentation by Li {\em et.~al}~\cite{li2019x8}.

%% file: conclusion.tex
Many of the state-of-the art methods for CPS design and implementation are not
model-based or involve a focus on individual model types --- for example, the
development of software synthesis techniques for specific models of computation
or reconfigurable architectures based on specific models for self-awareness and
adaptivity.  The study of design methodologies based on cooperating compact
system-level modeling approaches is a broad area that is ripe for further
study.  For example, deeper understanding is needed for many modeling
techniques on how these models may be adapted or parameterized to provide more
flexible trade-offs between model compactness and accuracy.  Some compact
modeling adaptations, such as hierarchical and factored MDPs~\cite{jons2006x1,
bout1995x1} or the multirate
versus homogeneous synchronous dataflow models of computation~\cite{lee1987x1}
(to name just a few), are established in the literature but are not applied in
practice to their full potential.  More diverse families of compact models,
more sophisticated design tool support for applying and integrating them, and
more concrete ways to assess the novel trade-offs introduced by such models are
all representative directions for future research that can help to address the
complexities and opportunities presented by heterogeneous CPS design.

%% file: ack.tex
We thank the following people who contributed to discussions about
heterogeneous computing and interoperability that have influenced this article:
Alvaro Cardenas, Roger Chamberlain, Tam Chantem, Changhee Jung, Miriam Leeser,
Shivakant Mishra, Mahdi Nikdast, Massimiliano Pierobon, Aviral Shrivastava,
Heechul Yun, and Ting Zhu.  This work was supported in part by the
U.S.~National Science Foundation under Grants CNS1514425 and CNS151304,
and by the U.S.~Air Force Office of Scientific Research under
Grant FA9550-18-1-0068.